# Observations of asteroid magnitude-phase relations at the Kharadze Abastumani Astrophysical Observatory

V. G. Shevchenko[1,2], R. Ya. Inasaridze[3,4], Yu. N. Krugly[1], V. V. Ayvazian[3,4], G. V. Kapanadze[3,4], G. Datashvili[3,4], I. G. Slyusarev[1,2], V. G. Chiorny[1], I. E. Molotov[5]

[1]Institute of Astronomy of V. N. Karazin Kharkiv National University, Kharkiv, Ukraine
[2]Department of Astronomy and Space Informatics of V. N. Karazin Kharkiv National University, Kharkiv, Ukraine
[3]Kharadze Abastumani Astrophysical Observatory, Ilia State University, K. Cholokoshvili Av. 3/5, Tbilisi 0162, Georgia
[4]Samtskhe-Javakheti State University, Rustaveli Street 113, Akhaltsikhe 0080, Georgia
[5]Keldysh Institute of Applied Mathematics, RAS, 4 Miusskaya sq., Moscow 125047, Russia

Email: shevchenko@astron.kharkov.ua

**Abstract**

In frame of an implementation of the cooperative program studying of asteroids between the Kharadze Abastumani Astrophysical Observatory and the Astronomical Institute of V.N. Karazin Kharkiv National University the observations for five main-belt asteroids were performed to obtain their magnitude-phase relations and other physical characteristics. Preliminary results of the photometrical observations for the large dark asteroid (1390) Abastumani are presented.

## 1. Introduction

The magnitude-phase relation is one of the main physical characteristics of asteroids obtained from observations. It allows first of all obtaining the absolute magnitude of asteroid, which on the one hand, is the basic for a determination of the albedo and diameter, and with another - the basic for ephemerid computations of apparent magnitudes. The magnitude-phase relation contains an information about physical properties of the asteroid's surface layer (i.e. light scattering mechanisms, albedo, roughness, porosity and refractive index of material). It is also used to determine a value of the phase integral for calculating albedo and diameters of asteroids from the emissivity in the infrared wavelength region (Masiero et al. 2011; Usui et al. 2011). All three areas differ for asteroids with different reflectivity and are formed in different scale structure of the surface: microrelief (scale up to hundreds of microns), mesorelief (up to tens of centimeters) and macrorelief (up to tenths of the average radius of an asteroid). The magnitude-phase dependence can be conventionally divided into three sections: the region of the opposition effect which usually shows the magnitude surge at the smallest phase angles (0÷7 deg), the linear region (7÷80 deg) and the nonlinear decrease region of brightness at largest phase angles (80÷180 deg). For the main belt asteroids, whose phase angles are typically less than 25-30 deg, the magnitude-phase dependence is linear down to 5-7 deg (described with linear phase coefficient), and has the opposition effect (OE), that is nonlinear brightness increasing near opposition, usually at phase angles less than 5-7 deg. The asteroids with different albedo show distinct differences in the linear slopes and the OE amplitudes (Belskaya, Shevchenko 2000). Most of asteroids show evident of the OE, but the some of low albedo asteroids do not show the OE (Shevchenko et al. 2012, 2014b; Slyusarev et al. 2014). A lack of the OE in asteroid phase dependences may indicates that these asteroids are the darkest objects in the Solar System. There is a relationship of the linear phase coefficient with geometric albedo (Belskaya, Shevchenko 2000, 2018). This relationship is very important because it gives a possibility to evaluate the asteroid albedo using photometric measurements only.

Practically all observations of asteroids are carried out at nonzero phase angles and to obtain their absolute magnitudes it is necessary to use a specially function. This function (i.e. a magnitude-phase function) can be also serve for the ephemeris computation of the apparent magnitude. Incorrect values of absolute magnitudes result in wrong values of albedos obtained from data of infrared surveys (WISE, AKARI, IRAS, etc.), that were demonstrated by Pravec et al. (2012) and Shevchenko et al. (2014a). Up to now, the two parameter *HG*-function was used for determination of an asteroid absolute magnitude (Bowell et al. 1989). But this function fits poorly the magnitude-phase relations of high and low albedo asteroids (Belskaya, Shevchenko 2000; Shevchenko et al. 2008, 2012). Recently new three parameter $HG_1G_2$-function has been proposed and actively used (Muinonen et al. 2010; Pentilla et al. 2016; Shevchenko et al. 2016). High quality data on magnitude-phase relations in the wide phase angle range were obtained for less than hundred main-belt asteroids. New high quality magnitude-phase relations for a large number of asteroids of different taxonomic classes are needed to perform a thorough check of this function.

Preliminary asteroid taxonomical classification can also be made using their magnitude-phase dependences. By measuring the slope of magnitude-phase dependence, it is possible to distinguish between low, moderate and high albedo surfaces (Belskaya, Shevchenko 2000; Carbognani et al. 2019; Mahlke et al. 2021; Oszkiewicz et al. 2012, 2021; Shevchenko et al. 2021). The main asteroid compositional types are well distinguished in the relationship of the OE amplitude and the linear slope (see Fig. 1).

This article presents new observations of several main belt asteroids, obtained in a wide range of phase angles and aimed at determining their magnitude-phase relations.

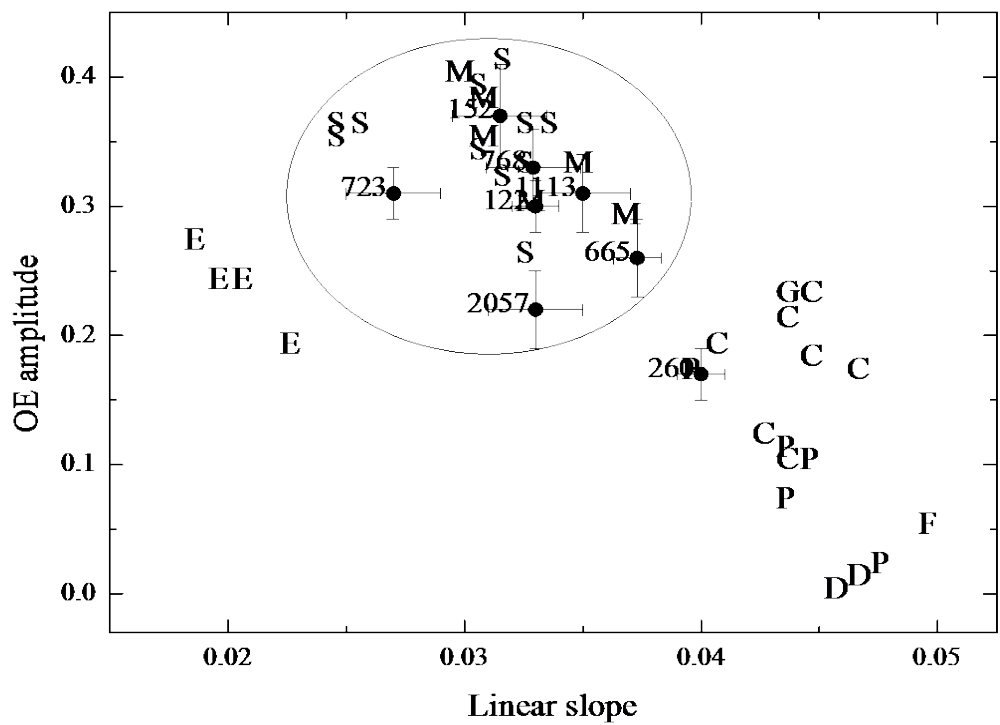

Fig. 1. Diagram of the OE amplitudes versus the linear slopes for asteroids of the main spectral clases (data taken from Belskaya, Shevchenko 2000; Shevchenko et al. 2021).

## 2. Observations and Results

The observations of magnitude-phase relations were launched at the Kharadze Abastumani Astrophysical Observatory in a frame of the cooperation with the Institute of Astronomy of V.N. Karazin Kharkiv National University. The observations of magnitude-phase relations of the selected asteroids were performed in the several directions: a) the study of the opposition effect of lowalbedo asteroids; b) the investigation of the phase curves of the V-asteroids; and c) the observations of the M-asteroids. The cooperative observations were performed from October 2018 to February 2021 for five asteroids during 62 nights. The physical characteristics of the observed asteroids (spectral class taken from Tholen (1989); albedo and diameter from Masiero et al. (2011, 2012) and Usui et al. (2011); and obtained in this work rotation period, color indices *B-V* and *V-R*, and absolute magnitude) are listed in the Table 1.

The CCD-observations of these asteroids were carried with the 70 cm Maksutov meniscus telescope AC-32 at the Kharadze Abastumani Astrophysical Observatory. The telescope is equipped with a FLI PL4240 Peltier-cooled CCD camera (2048×2048 pixels with size 13.5×13.5 μm). At the Institute of Astronomy of V.N. Karazin Kharkiv National University the CCD-observations were performed at Chuguev Observation Station using the 0.7-m reflector AZT-8. The telescope is equipped with a FLI ML4710 Peltier-cooled CCD camera (1056×1027 pixels with size 13×13 μm). Images were taken in *B*, *V* and *R* bands of the Johnson–Cousins photometric system. Original CCD-frames were reduced for dark current and flat field in the standard manner. The CCD-observations and data reduction methods were explained by Krugly et al. (2002, 2016). The brightness measurements of stars on CCD-images were done using the aperture photometry package ASTPHOT developed by Mottola et al. (1995). The absolute calibrations of the magnitudes were performed with standard star sequences taken from Landolt (1992) and Skiff (2007). In some cases for the calibration of the comparison stars we used their magnitudes in the SDSS (*ugriz*) photometric system, taken from the APASS DR9 (Henden et al. 2012) and Pan-STARRS DR1 (Chambers et al. 2016) catalogs. To transform them to the Johnson-Cousins (*UBVRI*) photometric system we used the corresponding equations given in Fukugita et al. (1996), and in Tonry et al. (2012). The accuracy of the resultant absolute photometry is within 0.01-0.03 mag. In next section, we present obtained results of the photometrical observations for the large dark main-belt asteroid (1390) Abastumani.

Table 1. Physical characteristics of the observed asteroids

| Asteroid | Sp. class | $p_v$ | D km | P, hours | B-V mag | V-R mag | H mag |
|---|---|---|---|---|---|---|---|
| (439) Ohio | P | 0.037 | 75.6 | 37.489±0.005 | 0.72 ± 0.02 | 0.41 ± 0.02 | 10.03± 0.02 |
| (863) Benkoela | A | 0.44 | 31.5 | - | 1.06 ± 0.04 | 0.56 ± 0.03 | 9.16± 0.02 |
| (1390) Abastumani | P | 0.033 | 98.3 | 13.164±0.001 | 0.71 ± 0.02 | 0.40 ± 0.02 | 9.42± 0.02 |
| (2263) Shaanxi | M | 0.16 | 22.3 | - | 0.81 ± 0.02 | 0.42 ± 0.02 | 12.20± 0.02 |
| (2763) Jeans | V | 0.41 | 7.5 | 7.80±0.02 | 0.82 ± 0.02 | 0.49 ± 0.02 | 12.38± 0.02 |

**2.1. (1390) Abastumani**

he asteroid was discovered on 3 October 1935 by Soviet astronomer Pelageya Shajn at the Simeiz Observatory in the Crimea. It was named after the Abastumani Astrophysical Observatory began working in Georgia in 1932. The asteroid is located in the outer part of the main belt with orbital parmeters: semi-major axis of 3.438 AU, eccentricity of 0.03 and inclination of 20 deg. The asteroid has large size of about 100 km and low albedo of the surface equal to 0.033 (Masiero et al. 2011; Usui et al. 2011), and it was classified as a rare P-type asteroid (Tholen 1989). An estimation of the rotation period (17.1 hour) was obtained by Gross (2003) using observations in April 2002. Durech et al. (2018) using the Lowell Photometric Database reconstructed shape model of this asteroid and obtained the rotation period equal to 13.16482 hours. Also there are estimations of the absolute magnitude *H* from Minor Planet Center (MPC) to be 9.19 mag and those obtained by Veres et al. (2015) (*H*=9.15) from Pan-STARRS survey.

Our observations of this asteroid were curried out in October-December 2018 and January 2019 for twelve nights in *B*, *V* and *R* bands of the Johnson–Cousins photometric system. The aspect data of asteroid are presented in the Table 2. The columns present the date of observation, ecliptic coordinates at epoch 2000.0, the distances from the asteroid to the Sun and to the Earth in astronomical unit (AU), the phase angle, the reduced *R* magnitude corrected for distances from the Earth and the Sun and corresponding to the primary maxima of the asteroid lightcurves, and their errors. We have determined the rotation period to be 13.164 ± 0.001 that is close to those determined by Durech et al. (2018). The composite lightcurve constructed with this period is pictured in Fig. 2. The maximal amplitude of lightcurve is about 0.35 mag. As it is seen on the Fig. 2, the lightcurve amplitude changes with increasing of the phase angle. We obtained also the color indices *B-V* and *V-*

*R*, which are presented in the Table 1.

The magnitude-phase relation for the maximum brightness of (1390) Abastumani in the *R* band is shown in Fig. 2. The maximum of brightness was used to take into account brightness variations of the asteroid with rotation and to build correctly the magnitude-phase relation. For an estimation of the absolute magnitudes of the asteroid, we used the new $HG_1G_2$– magnitude system proposed by Muinonen et al. (2010), with some modifications presented by Penttila et al. (2016). For actual computations, the online calculator of the *H*, $G_1$ and $G_2$ photometric parameters (http://h152.it.helsinki.fi/HG1G2/) was used. The dashed line on the Fig. 3 indicates the approximation of the phase curve by $HG_1G_2$-function with the parameters: $H_R = 8.85 \pm 0.02$ mag, $G_1 = 0.97 \pm 0.06$, $G_2 = 0.00 \pm 0.05$. The obtained values of $G_1$ and $G_2$ parameters are close to average values for the low albedo asteroids of D spectral class (Shevchenko et al. 2016). The magnitude-phase relation does not show an oppositional brightening and is linear in all range of observed phase angles with the linear phase coefficient equal to $0.046 \pm 0.001$ mag/deg. It points out very dark surface of this asteroid.

Table 2. Aspect data and measured magnitudes of asteroid (1390) Abastumani

| UT Date | $\lambda_{2000}$ | $\beta_{2000}$ | r | $\Delta$ | $\alpha$ | $R_o(1,\alpha)$ | |
|---|---|---|---|---|---|---|---|
| day | deg | deg | AU | AU | deg | mag | |
| 2018 10 05.81 | 32.423 | -1.244 | 3.327 | 2.371 | 5.96 | 9.147 | 0.022 |
| 2018 10 12.81 | 31.119 | -0.699 | 3.328 | 2.346 | 3.57 | 9.043 | 0.019 |
| 2018 10 14.87 | 30.717 | -0.537 | 3.329 | 2.341 | 2.84 | 9.016 | 0.018 |
| 2018 10 18.04 | 30.082 | -0.283 | 3.329 | 2.336 | 1.72 | 8.987 | 0.018 |
| 2018 10 19.03 | 29.882 | -0.204 | 3.329 | 2.335 | 1.36 | 8.947 | 0.019 |
| 2018 10 20.84 | 29.517 | -0.060 | 3.330 | 2.335 | 0.71 | 8.909 | 0.020 |
| 2018 10 29.69 | 27.730 | 0.647 | 3.331 | 2.345 | 2.45 | 9.020 | 0.020 |
| 2018 11 14.69 | 24.868 | 1.860 | 3.334 | 2.423 | 7.79 | 9.220 | 0.019 |
| 2019 01 12.70 | 24.626 | 4.892 | 3.345 | 3.151 | 17.08 | 9.688 | 0.020 |
| 2019 01 21.69 | 26.052 | 5.189 | 3.347 | 3.286 | 17.03 | 9.659 | 0.029 |
| 2019 01 23.80 | 26.428 | 5.254 | 3.348 | 3.318 | 16.98 | 9.677 | 0.020 |
| 2019 01 26.72 | 26.978 | 5.344 | 3.348 | 3.361 | 16.88 | 9.667 | 0.020 |

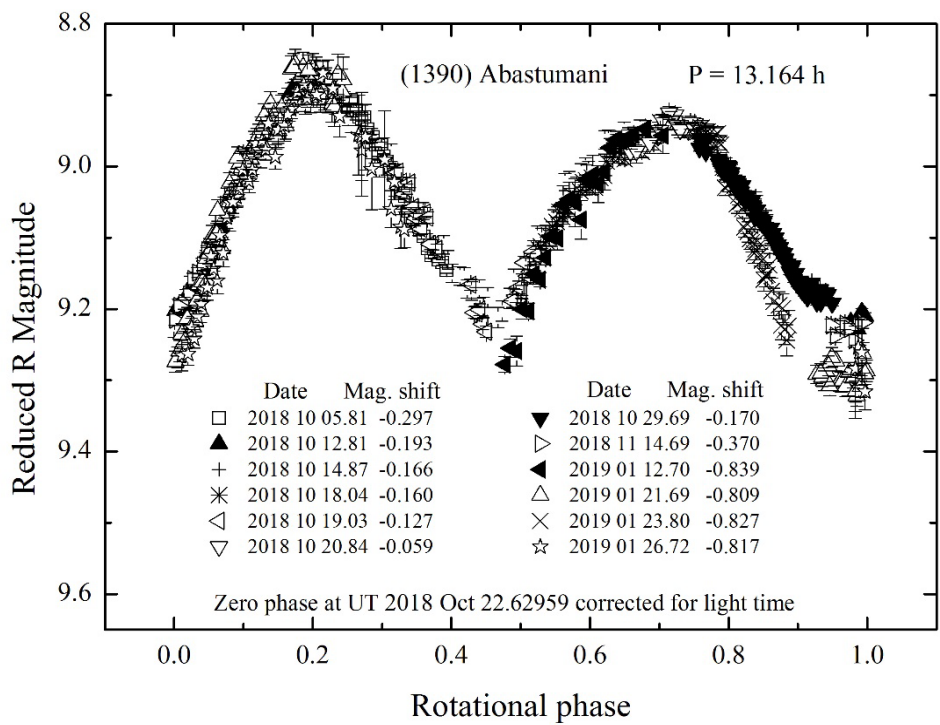

Fig. 2. Composite lightcurve of asteroid (1390) Abastumani

Taking into account the lightcurve amplitude and color index $V-R = 0.40$ mag, our estimate of the absolute magnitude in $V$ band is 9.42 mag. It should be noted that our estimation of the absolute magnitude has a difference with that given by the MPC ($H$=9.19) and to those obtained by Veres et al. (2015) ($H$=9.15).

## 3. Conclusions

As a result of an implementation of the proposed cooperative program we performed the photometric observations for five asteroids during 62 nights. The some main physical characteristics of the observed asteroids (rotation periods, color indices $B-V$ and $V-R$, and absolute magnitudes) were obtained. Here, we presented preliminary results of the photometrical observations for the large dark asteroid (1390) Abastumani. We obtained the magnitude-phase relation for this asteroid, which did not show the oppositional brightening. This is another asteroid without oppositional effect. Up to now, about ten of low-albedo asteroids without nonlinear increasing of brightness down to subdegree phase angles were detected among outer belt asteroids, Hildas and Jupiter Trojans (Shevchenko et al. 2014b; Slyusarev et al. 2014). These asteroids have a range of diameters from 50 to 200 km, and their magnitude-phase dependences are practically similar with small differences in linear slopes compared to uncertainties of measurements. All these asteroids belong mainly to the P and D spectral classes that have featureless spectra with moderate to high slope in the visual and near infrared wavelengths. It is necessary to obtain the spectrum of (1390) Abastumani for an unambiguous classification of this asteroid.

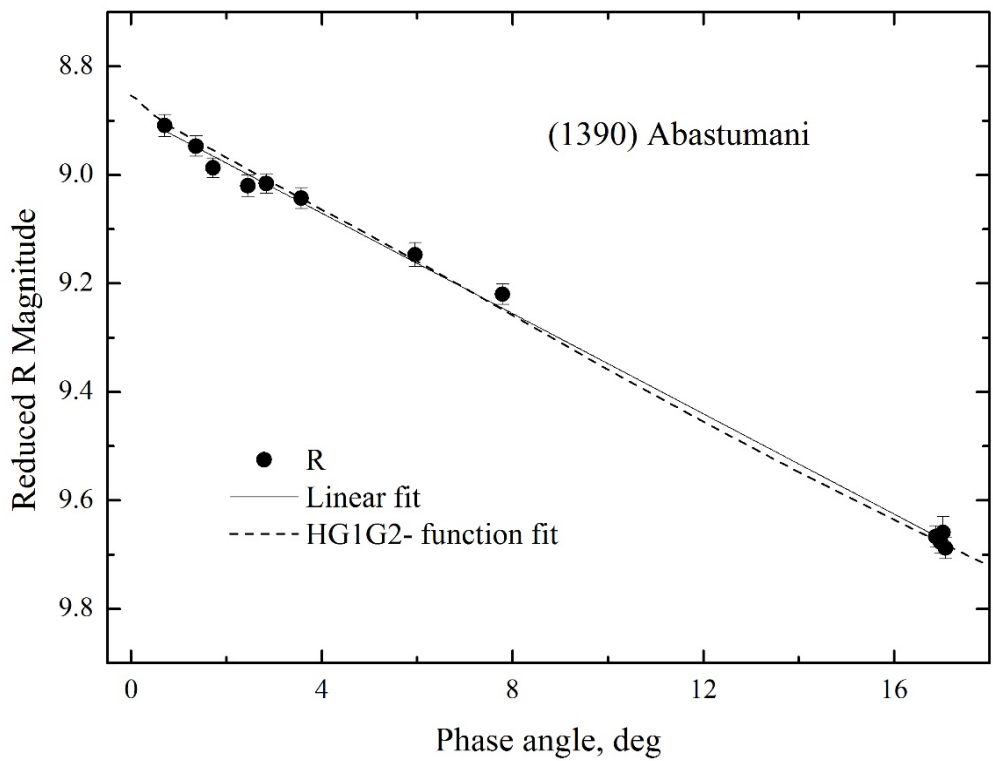

Fig. 3. Magnitude-phase relation of asteroid (1390) Abastumani

**Acknowledgments**

This research was partly supported by the Ukrainian Ministry of Education and Science. This work was partly prepared also as a result of the project, which was financed by the National Research

Foundation of Ukraine using the state budget: #2020.02/0371 "Metallic asteroids: search for parent bodies of iron meteorites, sources of extraterrestrial resources". Observations at Abastumani were partly supported by the Shota Rustaveli National Science Foundation, Grant RF-18-1193.